\documentclass[aps,prd,superscriptaddress,long,notitlepage,balancelastpage,nofootinbib,floatfix,twocolumn]{revtex4-1}
\pdfoutput=1

\usepackage{amsmath,mathtools,amssymb,amsbsy,amsthm,amsxtra,overpic,bbm,epsfig,subfigure,url}
\usepackage{mathrsfs}
\usepackage{color}
\usepackage{comment}
\usepackage{float}
\usepackage{enumitem}
\usepackage{slashed}
\usepackage{tikz}

\definecolor{nicered}{rgb}{0.5,0.,0.}
\definecolor{nicegreen}{rgb}{0.,0.5,0.}
\definecolor{niceblue}{rgb}{0.,0.,0.5}

\usepackage[colorlinks=true]{hyperref}
\hypersetup{
	linkcolor=niceblue,
	filecolor=nicegreen,      
	urlcolor=niceblue,
	citecolor=nicered,
}

\begin{document}
\title{Neutrino meets ultralight dark matter: $\boldsymbol{0\nu\beta\beta}$ decay and cosmology}
\author{Guo-yuan Huang}
\email{guoyuan.huang@mpi-hd.mpg.de}
\affiliation{Max-Planck-Institut f{\"u}r Kernphysik, Saupfercheckweg 1, 69117 Heidelberg, Germany} 

\author{Newton Nath} 
\email{newton.nath@ba.infn.it}
\affiliation{Istituto Nazionale di Fisica Nucleare,   Via  Orabona  4,  70126  Bari, Italy}

\begin{abstract}
\noindent
We explore the neutrinoless double beta ($0\nu \beta\beta$) decay induced by an ultralight dark matter field coupled to neutrinos.  
The effect on $0\nu\beta\beta$ decay is significant if the coupling violates the lepton number, for which the $\Delta L=2$ transition is directly driven by the dark matter field without further suppression of small neutrino masses.
As the ultralight dark matter can be well described by a classical field, the effect features a periodic modulation pattern in decay events.
However, we find that in the early Universe such coupling will be very likely to alter the standard cosmological results. 
In particular, the requirement of neutrino free-streaming before the matter-radiation equality severely constrains the parameter space, such that
the future $0\nu \beta\beta$ decay experiments can hardly see any signals even with a meV sensitivity to the effective neutrino mass.
\end{abstract}

\maketitle

\section{ Introduction}
\noindent
In spite of its remarkable success, the Standard Model (SM) of particle physics fails to address several fundamental issues, e.g., the nature of dark matter (DM).
The mass of  viable DM candidates can cover a huge landscape spanning from the primordial  black  holes~\cite{Bird:2016dcv,Carr:2016drx,Chen:2016pud,Georg:2017mqk} to the ultralight regime~\cite{Hu:2000ke}.
At present, the DM direct detection experiments such as Xenon1T \cite{XENON:2020gfr}, PandaX-II \cite{PandaX-II:2017hlx} come out with null signal of weakly interacting massive particle (WIMP), one of the most promising DM candidates, hence setting stringent limits on the available WIMP parameter space. 
Another  particular DM class, the ultralight DM~\cite{Ferreira:2020fam,Urena-Lopez:2019kud} with a mass ranging from $\mathcal{O}(10^{-22})~{\rm eV}$ 
\footnote{This is often called fuzzy dark matter~\cite{Hu:2000ke,Hui:2016ltb}. The term ``fuzzy'' corresponds  to a huge Compton wavelength, $\lambda = 2\pi /m_\phi \simeq 0.4\, {\rm pc} \times (10^{-22}\,\text{eV} / m_\phi)$ with $ m_\phi$ being the DM mass. }
to $\mathcal{O}(1)~{\rm eV}$, for example the well-motivated quantum chromodynamics (QCD) axion~\cite{Kim:2008hd,Graham:2015ouw,Irastorza:2018dyq,Marsh:2015xka}, is receiving renewing attention.
Different from WIMP, the ultralight DM is produced non-thermally in the early Universe and can be well described by a classical-number field.

Beside DM, the nature of massive neutrinos, i.e., whether they are Majorana or Dirac fermions, is yet to be answered. At present, the only feasible process that can uncover the nature of neutrinos is the neutrinoless double beta ($ 0\nu\beta \beta $) decay,  which violates the lepton number by two units~\cite{DellOro:2016tmg,Vergados:2016hso}.

There have been many attempts to connect the mentioned two mysterious sectors, i.e., neutrinos and ultralight DM~\cite{Berlin:2016woy,Brdar:2017kbt,Krnjaic:2017zlz,Liao:2018byh,Capozzi:2018bps,Reynoso:2016hjr,Huang:2018cwo,Pandey:2018wvh,Farzan:2018pnk,Choi:2019ixb,Baek:2019wdn,Choi:2019zxy,Choi:2020ydp,Dev:2020kgz,Baek:2020ovw,Losada:2021bxx,Smirnov:2021zgn,Alonso-Alvarez:2021pgy,Huang:2021zzz}. 
In this work, we investigate $0\nu\beta\beta$ decays in the presence of a coupling between neutrinos and  the ultralight DM with a general mass spanning from $10^{-22}~{\rm eV}$ to sub-eV.
Among general interaction forms, the scalar and tensor couplings are able to violate the lepton number (or equivalently, the DM particle carries two units of lepton number), and hence important for  $0\nu\beta\beta$-decay experiments. The Lagrangian reads
\begin{eqnarray}\label{eq:NU-DM-Inter}
-\mathcal{L}^{\rm M}_{\rm int} = 
 \begin{dcases} 
\frac{g_{\phi}^{\alpha \beta}}{2} \, \phi \, \overline{\nu^{}_{\rm L \alpha}}  \nu^{\rm c}_{\rm L \beta} + {\rm h.c.}\;,  \\[0.5ex]
\frac{g_{T}^{\alpha \beta}}{2} T^{\mu \nu}_{} \overline{\nu^{}_{\rm L \alpha}}  \sigma^{}_{\mu\nu} \nu^{\rm c}_{\rm L \beta}  + {\rm h.c.} \;,
\end{dcases}
\end{eqnarray}
where $\alpha, \beta = e, \mu,\tau$ are the flavor indices, and $\phi$ and $T^{\mu \nu}$ represent real scalar and tensor fields with neutrino coupling constants $g_{\phi}^{\alpha \beta}$ and  $g_{T}^{\alpha \beta}$, respectively. 
We have assumed that neutrinos are Majorana particles, such that only the field $\nu^{}_{\rm L}$ is involved at low energies.

In analogy with the electromagnetic force, the tensor interaction can be generated by a dark photon field $A'$ coupled to neutrinos with a dark magnetic dipole moment~\cite{Capozzi:2018bps}. However, for Majorana neutrinos, only transition magnetic moment exists and the diagonal part of the tensor coupling constant $g_{T}^{\alpha \beta}$ is vanishing, thus not contributing to the $0\nu\beta\beta$ process which exchanges two electron neutrinos.
Hence, we will restrict our following discussions to only the ultralight scalar DM.
It is worth mentioning that the transition magnetic moment can be relevant for the lepton-number-violating meson decays by exchanging different flavors of neutrinos.

For Majorana neutrinos, $\nu \equiv \nu^{}_{\rm L} + {\nu^{\rm c}_{\rm L}}$, in the background of scalar DM, the Hamiltonian density of the system is
\begin{eqnarray}
\mathcal{H} = \frac{1}{2} (\partial \phi)^2 + \frac{1}{2} m^{2}_{\phi} \phi^2 + \frac{1}{2}\overline{\nu} (\mathrm{i} \slashed{\partial} + \widehat{M}^{}_{\nu} + g \phi) \nu \;,
\end{eqnarray}
where $m^{}_{\phi}$ is the DM mass, $\widehat{M}^{}_{\nu}$ is the neutrino mass matrix in vacuum, and the coupling constant $g$ is a real symmetric matrix in general. Note that the pseudoscalar coupling has been ignored in the expression, which will induce a similar effect.
The corresponding equation of motion is given by
\begin{eqnarray}
& &(\partial^2 + m^2_{\phi})\phi = - \frac{1}{2}  \overline{\nu}^{}_{\alpha} g^{}_{\alpha\beta} \nu^{}_{\beta}\;, \label{eq:EOMphi} \\ 
& &i \slashed{\partial}  \nu^{}_{\alpha} = ({M}^{}_{\nu,\alpha\beta} + g^{}_{\alpha\beta} \phi) \nu^{}_{\beta} \;,\label{eq:EOMnu}
\end{eqnarray}
where the flavor indices are explicitly kept.
A proper diagonalization is required to obtain the evolution equation of neutrino mass eigenstates.
As long as the neutrino number density is not too large in Eq.~(\ref{eq:EOMphi}), the ultralight DM evolves as free classical field in the Universe as $\phi = \phi^{}_{0} \cos{m^{}_{\phi}t}$ after the scalar mass ``freeze-in'', i.e., $H < m^{}_{\phi}$, with $H$ being the Hubble expansion rate. The field strength is given by $\phi^{}_{0} =\sqrt{2\rho}/m^{}_{\phi}$, where $\rho \approx m^{}_{\phi} n^{}_{\phi} \approx 0.3~{\rm GeV \cdot cm^{-3}}$ is the DM energy density in our local galaxy.

\begin{figure}[t!]
	\begin{center}
		\includegraphics[width=0.2\textwidth]{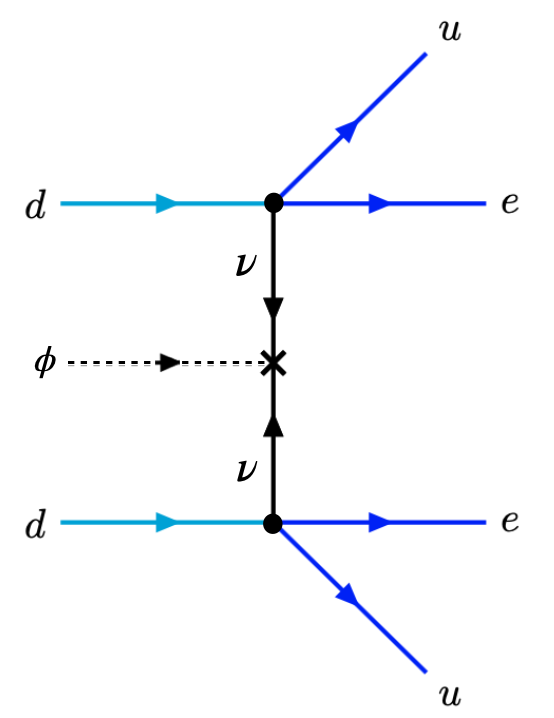}
		\hspace{0.3cm}
		\includegraphics[width=0.2\textwidth]{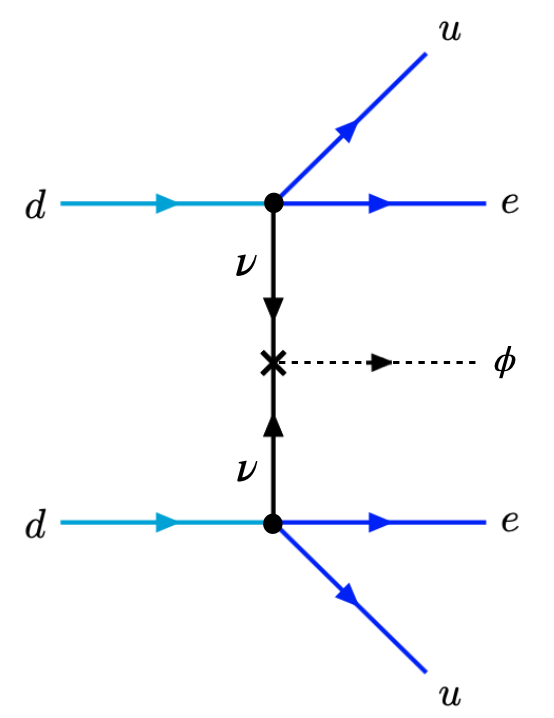}
	\end{center}
	\vspace{-0.3cm}
	\caption{\footnotesize Feynman diagrams of $0\nu\beta\beta$ decays induced by the absorption and emission of ultralight scalar DM. Because these transitions take place in a background of identical scalar bosons, there will be an enhancement of stimulation. }
	\label{fig:Feyn}
\end{figure}

The large occupation number (i.e., $N=n^{}_{\phi} \lambda^3 \approx 6 \times 10^{36} (10^{-10}{\rm eV}/m_{\phi})^4$) of local ultralight DM within the cube of Compton wavelength, guarantees the classical-field description as an excellent approximation to the more fundamental quantum field.
The formation of the classical field in our local galaxy will generate a time-varying effective Majorana mass $m^{}_{\nu, {\rm eff}} = g\phi$ to neutrinos, which is very similar to the Higgs mechanism.  
At the particle level, the $0\nu\beta\beta$ decays induced by ultralight DM are shown as  diagrams in Fig.~\ref{fig:Feyn}. The illustrated leading-order contribution is accurate enough since the coupling of our concern is set to be extremely small. 
If the ultralight DM is a vector, we will need an additional Majorana neutrino mass insertion, such that the final amplitude will be proportional to $gV^{}_{\mu} \cdot m^{}_{\nu}/Q$ instead of $g\phi$ for the scalar case, where $Q$ is the typical momentum transfer of $0\nu\beta\beta$ decays.
The emission process has been discussed before in the context of Majoron model~\cite{Burgess:1992dt,Burgess:1993xh,Blum:2018ljv,Brune:2018sab,Cepedello:2018zvr}.
However, different from the ``spontaneous'' Majoron emission process, in our case both the absorption and emission of scalars benefit from the large occupation number of ultralight DM, which gives rise to a dramatic stimulation enhancement.
In fact, the enhancement makes the coupling as small as $10^{-23}$ to be relevant as we will see later, while for the stochastic Majoron emission process this coupling leads to negligible decay rate.
Moreover, the energy difference of the absorption and emission processes is as tiny as the DM mass. This is negligible compared to the energy resolution of electrons from $0\nu\beta\beta$ decays, and hence the absorption and emission processes should be added coherently. We will find in later discussions that this is equivalent to treating the ultralight scalar DM as a classical field.

In the rest of the work, we first describe how the coupled system of ultralight DM and neutrinos evolves in the early Universe in Sec.~\ref{sec:cosmo}.
Based on the robust cosmic microwave background (CMB) observations, we confine the allowed parameter space of the mass and coupling of ultralight DM. 
In Sec.~\ref{sec:0NuBB}, we explore the possible impacts on $0\nu\beta\beta$ decays. Conclusions are presented in Sec. \ref{sec:conclusion}.

\section{Cosmological evolution} \label{sec:cosmo}
\noindent
A popular way to generate the coherent scalar field in the early Universe is via the misalignment mechanism, which is well known for the QCD axion~\cite{Kim:2008hd,Graham:2015ouw,Irastorza:2018dyq,Marsh:2015xka}.
The production is typically associated with a global symmetry, which is broken spontaneously and results in a Goldstone mode.
The later dynamics will break the shift-symmetry of the Goldstone by selecting a preferred phase. A mass term will then be generated to the field, which populates the DM energy density after $H \lesssim m^{}_{\phi}$.
The initial condition of a misaligned scalar in our scenario reads~\cite{Marsh:2015xka}
\begin{eqnarray}\label{eq:init}
\phi(t^{}_{\rm i},x) = f^{}_{\phi} \theta^{}_{\phi} \;,~\dot{\phi}(t^{}_{\rm i},x) = 0 \;,
\end{eqnarray}
where $t^{}_{\rm i }$ is the time set by the spontaneous symmetry breaking of the global symmetry, $f^{}_{\phi}$ represents the energy scale of the symmetry breaking, $\theta^{}_{\phi}$ is the initial phase, and $\dot{\phi} \equiv \mathrm{d}\phi/\mathrm{d}t$.

At the quantum field level, the ultralight DM produced by the misalignment mechanism (from the dynamical symmetry breaking) should be descried by a coherent state of the form~\cite{Marsh:2015xka,Ferreira:2020fam,Miransky1994,Ferreira:lecture,Davidson:2014hfa,Veltmaat2020GalaxyFW,Guth:2014hsa,Zhang:1999is,Barnett:1997}
\begin{eqnarray}
\left|\phi \right\rangle = \mathrm{e}^{-\frac{N}{2}} \mathrm{e}^{\int \frac{\mathrm{d}^3 k}{(2\pi)^3 } \tilde{\phi}(k) \hat{a}^\dagger_{k}}  \left|0 \right\rangle \;,
\label{eq:cstate}
\end{eqnarray}
where $\mathrm{e}^{-\frac{N}{2}}$ is a normalization factor, $\tilde{\phi}(k)$ is a general complex field with the wave-vector $k$, and $\hat{a}^\dagger_{k}$ is the corresponding particle creation operator. 
The coherent state  $| \phi \rangle$ has been thought of as the most field-like solution of quantum states.
The property of this state can be largely captured by the expectation value of the field operator $\hat{\phi}$, namely 
\begin{eqnarray}
\phi(t,x) \equiv \left\langle \phi \right| \hat{\phi} \left|\phi \right\rangle & = & \int \frac{\mathrm{d}^3 k}{(2\pi)^3 } \frac{1}{ \sqrt{2 E^{}_{k}}} \left[ \tilde{\phi}(k) \mathrm{e}^{-i (\omega t - k \cdot x)} + \right. \notag\\
& & \left. + \tilde{\phi}^*(k) \mathrm{e}^{i (\omega t - k \cdot x)}\right] \;,
\label{eq:}
\end{eqnarray}
which is a classical-number field. This non-vanishing  expectation value can be interpreted in a way similar to the Higgs mechanism.
The evolution of the classical field $\phi(t,x)$ right follows the equation of motion in Eq.~(\ref{eq:EOMphi}).

For the isotropic DM field without kinematic energy, i.e., $k=0$, we have $\left|\phi \right\rangle = \mathrm{e}^{-\frac{N}{2}} \mathrm{e}^{ \sqrt{N} \mathrm{e}^{\mathrm{i}\theta} \hat{a}^\dagger_{0}}  \left|0 \right\rangle$.
It should be emphasized that the coherent state $| \phi \rangle$ is an eigenstate of $\hat{a}^{}_{0}$, i.e., $\hat{a}^{}_{0} | \phi \rangle = \sqrt{N} \mathrm{e}^{\mathrm{i}\theta} | \phi \rangle$, but not that of the particle number $\hat{N} = \hat{a}^{\dagger}_{0}\hat{a}^{}_{0}$ in the Fock space.
However, for a large enough ensemble of DM, there is a well-defined mean particle number $\langle \hat{a}^{\dagger}_{0}\hat{a}^{}_{0} \rangle = N$ with a quantum fluctuation proportional to $1/\sqrt{N}$.
In fact, for the state with a definitive particle number, e.g., $\hat{a}^{\dagger}_{0}\hat{a}^{}_{0}| N \rangle = N | N \rangle$, the field expectation value $\left\langle N \right| \hat{\phi} \left|N \right\rangle$ is vanishing.
The difference between $| \phi \rangle$  and $| {N} \rangle$ is critical and should be highlighted, as they will induce very distinct signals experimentally.
For instance, for $|{N} \rangle$ the absolute quantum phase $\mathrm{exp}(-\mathrm{i}\, m^{}_{\phi}t)$ is meaningless in practice, but the classical phase of $| \phi \rangle$ is physical and will lead to time-varying signals~\cite{DirectMeasurementofLightWaves,Barnett1996}.

In the early Universe, the problem is ascribed to solving the evolution of classical field.
Because the ultralight DM should be homogeneously produced, the evolution equation of Eq.~(\ref{eq:EOMphi}) can be recast to~\cite{Esteban:2021ozz}
\begin{eqnarray}\label{eq:phiEvol}
\ddot{\phi}+3 H \dot{\phi} + m^2_{\phi} \phi = - g \int \frac{\mathrm{d}^3 p}{(2\pi)^3} \frac{g \phi}{\sqrt{p^2+(g\phi)^2}} f^{}_{\nu}(a|p|) , \hspace{0.5cm}
\end{eqnarray}
where $a$ is the scale factor, connected to the redshift $z$ via $a \equiv a^{}_{0}/(z+1)$ with $a^{}_{0}$ being the scale factor at present. 
Here, the neutrino distribution function simply follows $f^{}_{\nu}(a|p|) = D^{}_{\nu}/(\mathrm{e}^{|p| / T^{}_{\nu}(a)} + 1 )$ with the temperature $T^{}_{\nu}(a) \propto 1/a$ and the factor $D^{}_{\nu}=6$ accommodating three generations of neutrinos and antineutrinos. 
Note that we require that the scalar acquires the shift-symmetry breaking terms, including its mass and coupling to neutrinos, after the Big Bang nucleosynthesis (and hence also the neutrino decoupling), otherwise the primordial element abundances will be significantly affected~\cite{Venzor:2020ova}. Thus, the distribution function $f^{}_{\nu}(a|p|)$ is the same as that in the standard case without scalar couplings.
The source term on the right-hand side of Eq.~(\ref{eq:phiEvol}) modifies the evolution of a free scalar. Note that we have assumed the vacuum neutrino mass to be vanishing in the above equation, and a larger effect of neutrino source will be expected if the vacuum neutrino mass is taken into account.
To simplify the analysis of cosmological effect, we have assumed the coupling matrix $g$ to be universal for all three neutrino flavors. The flavor indices are not given explicitly here, and $g$ will represent the general order of magnitude of the coupling matrix elements.
We comment on the effect if the elements of $g$ are non-diagonal and hierarchical.
In fact, because three flavors of neutrinos have a similar distribution function, the right-hand side of Eq.~(\ref{eq:phiEvol}) will always be dominated by the largest possible elements. For instance, suppose only the element $g^{}_{e\mu}$ ($= g^{}_{\mu e}$) to be non-zero, then we will have a similar source term as Eq.~(\ref{eq:phiEvol}) but with $g$
replaced by $g^{}_{e\mu}$, and the degrees of freedom are also reduced to $D^{}_{\nu}=2$.

\subsection{Observable impacts}
\noindent
If $T^{}_{\nu} \gg g\phi$ is satisfied (i.e., neutrinos in the plasma are relativistic), the source term is equivalent to a screening mass of the scalar, with~\cite{Davoudiasl:2018hjw,Esteban:2021ozz}
\begin{eqnarray}\label{eq:}
m^2_{\phi,\rm s}(a) = \int \frac{\mathrm{d}^3 p}{(2\pi)^3} \frac{g^2 }{|p|} f^{}_{\nu}(a|p|) \sim g^2 T^2_{\nu} . \hspace{0.5cm}
\end{eqnarray}
The screening mass describes how the neutrino plasma should response to the change of the scalar potential, and its main effect will be altering the oscillating frequency of scalar field. In this case, we have
\begin{eqnarray}\label{eq:evolphi}
\ddot{\phi}+3 H \dot{\phi} + \left[m^2_{\phi}+ m^2_{\phi,\rm s}(a) \right] \phi = 0 \;, \hspace{0.5cm}
\end{eqnarray}
with an effective mass varying as the Universe expands, $m^{}_{\phi,{\rm eff}} = [m^2_{\phi}+ m^2_{\phi,\rm s}(a) ]^{1/2}$. 
Note that the scalar particle produced from misalignment mechanism is non-relativistic (hence cold) with almost vanishing momentum. Because the screening mass is expected to be homogeneous in the Universe, the scalar momentum will not be changed under the screening effect according to the translational symmetry, and only its energy will be affected.
For the numerical convenience, it is useful to replace the physical time $t$ with the scale factor $a = m\, x$, where $m$ is an arbitrary scale and $x$ is dimensionless. The equation of motion can be recast into
\begin{eqnarray}\label{eq:phiprime}
x^2 \phi^{\prime\prime} + \left( 4 x + x^2 \frac{H^\prime}{H} \right) \phi^{\prime} + \frac{m^2_{\phi,{\rm eff}}}{H^2} \phi = 0 \;,
\end{eqnarray}
where ${\phi}^{\prime} \equiv \mathrm{d}\phi/\mathrm{d}x$.

The scalar field $\phi$ starts to oscillate when $H \lesssim m^{}_{\phi,{\rm eff}}$. 
If $\phi$ is not coupled to neutrinos, its amplitude will simply scale as $\phi^{}_{0} \propto a^{-3/2}$~\cite{Marsh:2015xka}, and the energy density $\rho_{\phi} \approx m^{2}_{\phi} \phi^2_{0}/2$ will be proportional to $a^{-3}$, which justifies its capability of being the cold DM.
However, if the scalar mass is significantly screened, this benefit will be lost due to the behavior $m^{}_{\phi,{\rm s}} \propto a^{-1}$. In the completely screened case, i.e. $m^{}_{\phi,{\rm s}} \gg m^{}_{\phi}$, we find
\footnote{In the fast-oscillating regime, to obtain the evolution of the amplitude we decompose the scalar field as $\phi(t) = \phi^{}_{0}(t) \cos{(m^{}_{\phi,{\rm eff}} \cdot t)}$~\cite{Marsh:2015xka}. Keeping the leading order in the assumption of $H \sim \dot{\phi}^{}_{0} \ll m^{}_{\phi,{\rm eff}} $, we obtain $2 \dot{\phi^{}_{0}} = - 3 H \phi^{}_{0} - \phi^{}_{0} \dot{m}^{}_{\phi,{\rm eff}}/m^{}_{\phi,{\rm eff}} $. For $m^{}_{\phi,{\rm eff}} \propto a^{-1}_{}$, we have $\dot{\phi^{}_{0}} = -H \phi^{}_{0}$, and hence $\phi^{}_{0} \propto a^{-1}$. }
\begin{eqnarray}\label{eq:}
\phi^{}_{0} \propto a^{-1} \;,~ \rho^{}_{\phi} \sim  \frac{m^{2}_{\phi,{\rm s}} \phi^2_{0}}{2} \propto  a^{-4} \;,
\end{eqnarray}
which indicates that the energy of $\phi$ evolves more like a {``radiation''}  with the screened mass instead of the ordinary cold DM.
One can also have an intuitive understanding of the result from the particle level.
Since the particle number here is a conserved quantum number, the scalar number density simply follows the relation $n^{}_{\phi} \propto a^{-3}$. 
If the screening mass is dominant (i.e., $m^{}_{\phi,{\rm s}} \gg m^{}_{\phi}$), 
we arrive at $\rho^{}_{\phi} \approx m^{}_{\phi,{\rm s}} n^{}_{\phi} \propto a^{-4}$, because $m^{}_{\phi,{\rm s}} \propto a^{-1}$ is fixed by the neutrino temperature. 
The scalar energy density evolves like a ``radiation'' simply due to the dominance of the varying screening mass. 
We should emphasize that the DM perturbation evolution will still follow that of the ordinary cold DM, because our scalar particles are always non-relativistic. Hence, the structure formation will not be affected as in the case of hot DM scenario, where small structures are smoothed out due to the large DM velocity.
The observable difference made by the screening effect is to change the background evolution, e.g., delay the matter-radiation equality.
The screening effect is thus to be constrained by the CMB observations if we require the scalar energy evolution to follow the ordinary DM.
Recall that in our scenario, we require the scalar to be a dynamical DM field with the initial condition in Eq.~(\ref{eq:init}).
In contrast, Ref.~\cite{Esteban:2021ozz} has discussed the scenario where $\phi$ is only sourced by neutrinos but not a dynamical field.

\begin{figure}[t!]
	\begin{center}
		\includegraphics[width=0.43\textwidth]{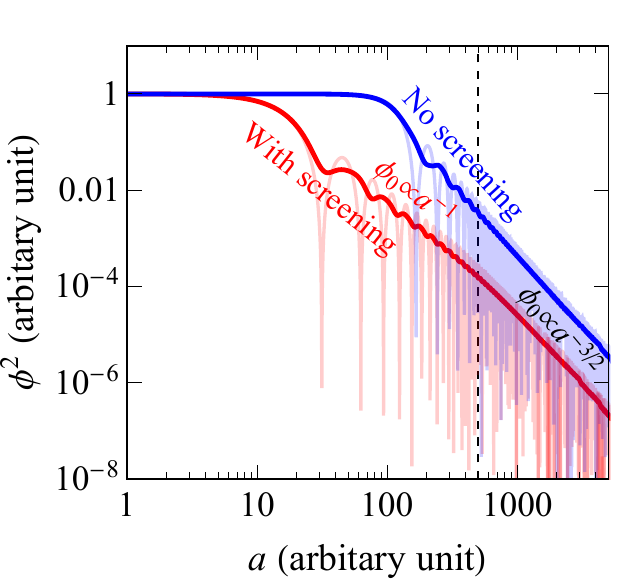}
	\end{center}
	\vspace{-0.3cm}
	\caption{\footnotesize A demonstration of evolution of ultralight DM as a function of the scale factor. For demonstration we give $\phi^2$ in the logarithmic scale.  The case with the screening effect of neutrinos is shown as the red curve, while that without the screening effect is given as the blue one. The lighter curves are the actual solutions, and the darker ones are averaged over certain scale factor to see the magnitude of the field.}
	\label{fig:phi}
\end{figure}

An example of $\phi^2$ ($\geq 0$) evolution with respect to the scale factor $a$ is given in Fig.~\ref{fig:phi} by numerically solving Eq.~(\ref{eq:phiprime}). 
The lighter fast-oscillating curves stand for the magnitude of $\phi^2$ with (red) or without (blue) the screening effect, and the darker ones are produced by averaging over fast-oscillating modes to see the overall amplitude.
As the screening mass $m^{}_{\phi,{\rm s}}$ is dominant for $a<500$, the scalar field scales as $a^{-1}$ instead of $a^{-3/2}$ and contributes to the radiation component. After $a=500$ the vacuum mass starts to dominate over the screening one, and the scalar field plays the role of DM as normal.

Another effect to avoid is the impact on neutrino velocity $v^{}_{\nu}$, which is related to the neutrino free-streaming length before the decoupling of CMB.
As has been mentioned, in the background of ultralight DM, neutrinos acquire an effective mass  $m^{}_{\nu, {\rm eff}} = g \phi $. The neutrino velocity in the early Universe will be given by
\begin{eqnarray}\label{eq:vnu}
\frac{v^{}_{\nu}}{c} = \frac{|p|}{\sqrt{p^2+ m^{2}_{\nu, {\rm eff}}}} \;,
\end{eqnarray}
where the average momentum of neutrino plasma reads $\langle |p| \rangle \sim 3 T^{}_{\nu}$. 
Notice that the effective neutrino mass $m^{}_{\nu, {\rm eff}}\propto a^{-3/2}$ increases faster than $T^{}_{\nu}\propto a^{-1}$ as we go to higher redshift $z$. This means that neutrinos will become more and more non-relativistic and hence less and less free-streaming at higher $z$.
In comparison, the CMB data are consistent with neutrino free-streaming with a speed of light up to a redshift as high as $z=10^5$~\cite{Hannestad:2004qu, Hannestad:2005ex, Bell:2005dr, Basboll:2008fx, Archidiacono:2013dua, Forastieri:2015paa, Cyr-Racine:2013jua, Oldengott:2014qra, Forastieri:2017oma,Oldengott:2017fhy,Choudhury:2020tka,Brinckmann:2020bcn,Das:2020xke,Mazumdar:2020ibx}.
For simplicity, in the following we will consider neutrinos as free-streaming with the speed of light if $T^{}_{\nu} > g \phi$, and vise versa.

It is worth noting that the CMB constraint on the sum of neutrino masses is not as competitive as the above two effects. The combined analysis of Planck  TT, TE, EE + lowE + lensing + BAO data leads to $\Sigma m^{}_{\nu} < 0.12~{\rm eV}$~\cite{Planck:2018vyg}, which actually takes account of data  at very low redshift long after the recombination $z^{}_{\rm rc} = 1100$.
In fact, around the recombination era neutrinos with a mass $\mathcal{O}(0.1)~{\rm eV}$ are still relativistic.
Since $m^{}_{\nu, {\rm eff}} \propto a^{-3/2}$,
the impact of effective neutrino masses turns out to be less and less important at lower $z$.
Hence, we do not attempt to set the bound according to $\Sigma m^{}_{\nu,\rm eff}(z) < 0.12~{\rm eV}$ at recombination  $z^{}_{\rm rc} = 1100$, which has been considered previously.

\begin{figure}[t!]
	\begin{center}
		\includegraphics[width=0.4\textwidth]{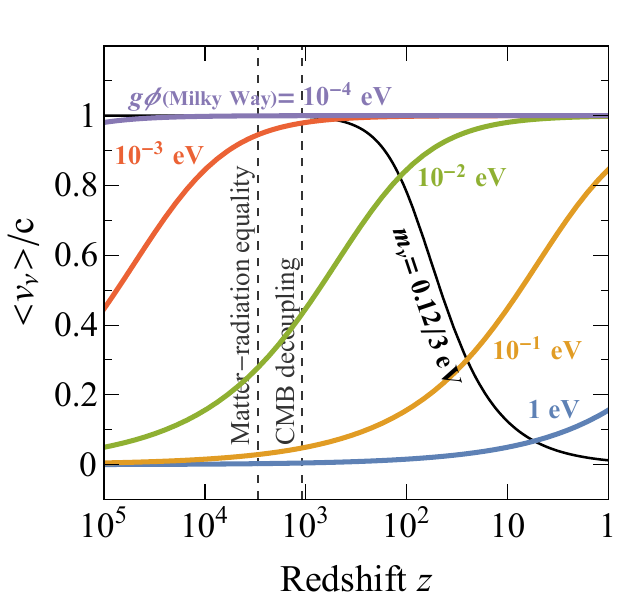}
	\end{center}
	\vspace{-0.3cm}
	\caption{\footnotesize The evolution of average neutrino velocity in terms of the redshift in the early Universe. The colorful curves stand for the cases with a coupling between neutrinos and the ultralight DM, with the vacuum neutrino mass vanishing and DM-induced neutrino mass $g\phi$ at present indicated close to the curve. The black curve represents the neutrino velocity of the standard case with a mass $m^{}_{\nu} = 0.04~{\rm eV}$.}
	\label{fig:vnu}
\end{figure}

\subsection{Updated CMB constraints}
\noindent Depending on the parameter choice of $g$ and $m^{}_{\phi}$, there can be four different scenarios in the early Universe for the evolution equation of Eq.~(\ref{eq:phiEvol}):
\begin{itemize}
	\item {\bf Case~I}: $ g \phi < T^{}_{\nu}$ and $g^2 T^2_{\nu} < m^2_{\phi}$. The effective neutrino mass is negligible compared to the neutrino temperature and the screening effect is also not important. In this case, the scalar field evolves like matter with $\phi^{}_{0} \propto a^{-3/2}$ while $T^{}_{\nu} \propto a^{-1}$, and these two conditions will maintain themselves as the Universe expands.
	\item {\bf Case~II}: $ g \phi < T^{}_{\nu}$ and $g^2 T^2_{\nu} > m^2_{\phi}$. Neutrinos are relativistic but the screening scalar mass $m^{}_{\phi,\rm s}\sim g T^{}_{\nu}$ dominates the oscillation of $\phi$.
	 In such a scenario, the energy density of the scalar field evolves like a ``radiation'' because of the screening mass. But, note again that the scalar particles are always non-relativistic.
	Furthermore, the induced neutrino effective mass $g \phi$ has a similar scaling relation to its temperature $T^{}_{\nu}$  with respect to the scale factor.
	Because the ratio of $T^{}_{\nu}$ to $g \phi$ measures the rapidity, neutrinos will remain to be relativistic if it is satisfied initially, as long as $g^2 T^2_{\nu} > m^2_{\phi}$. However, this case will eventually collapse into {\bf Case~I} after $g^2 T^2_{\nu} < m^2_{\phi}$.
	\item {\bf Case~III}: $g \phi > T^{}_{\nu}$ and $g T^3_{\nu} < m^2_{\phi} \phi$. The screening effect is not important compared to the vacuum mass of $\phi$, but neutrinos are non-relativistic (thus not free-streaming with $c$) due to the large effective mass. This case will also end up into {\bf Case~I} as the Universe expands.
	\item {\bf Case~IV}: $g \phi > T^{}_{\nu}$ and $g T^3_{\nu} > m^2_{\phi} \phi$. Both the screening effect and the effective neutrino mass are dominant factors, which is to be avoided over the observable history of CMB.
\end{itemize}
For {\bf Cases~I} and {\bf III}, in terms of redshift, the neutrino temperature and the scalar field strength read as $T^{}_{\nu} \approx 1.68 \times 10^{-4}~{\rm eV} (z+1)$ and $\phi \approx 2.15 \times 10^{-3}~{\rm eV}^2 / m^{}_{\phi} \left[(z+1)/45 \right]^{3/2}$, respectively, assuming the ultralight scalar accommodates all the DM abundance.

The evolution of the Universe can be in a sequence of {\bf Case~III}$\to${\bf I} or {\bf II}$\to${\bf I}.
In Fig.~\ref{fig:vnu}, we illustrate the neutrino velocity evolution $\langle v^{}_{\nu} \rangle/c $, i.e., Eq.~(\ref{eq:vnu}), in terms of the redshift $z$ from {\bf Case~III} to {\bf I}, where the screening effect is not important but the neutrino effective mass is severely affected.
This can be achieved by properly choosing the DM parameter.
It can be noticed that the neutrino velocity drops rapidly as we go back to higher $z$, which should be avoided to be in accordance with CMB free-streaming requirement.
In comparison, we also give the neutrino velocity in the standard case with a constant mass $m^{}_{\nu} = 0.04~{\rm eV}$ corresponding to the Planck limit $\Sigma m^{}_{\nu} < 0.12~{\rm eV}$. In this case, the neutrino velocity evolves in an opposite manner compared to that with a coupling to the ultralight DM. As has been mentioned, around the recombination era $z^{}_{\rm rc} \approx 1100$, neutrinos of the Planck limit are still ultra-relativistic, and it is the data at lower $z$ that contribute to this limit.

In order not to significantly alter the CMB observations, two criteria should be satisfied: (1) the ultralight scalar must evolve as matter before the matter-radiation equality around $z^{}_{\rm eq} = 3000$; (2) neutrinos should be free-streaming up to the redshift $z^{}_{\rm eq} = 3000 $ (conservative) or $z^{}_{\rm fs} = 10^5$ (aggressive). The aggressive redshift for neutrino free-streaming is inspired by the investigation of neutrino self-interactions~\cite{Hannestad:2004qu, Hannestad:2005ex, Bell:2005dr, Basboll:2008fx, Archidiacono:2013dua, Forastieri:2015paa, Cyr-Racine:2013jua, Oldengott:2014qra, Forastieri:2017oma,Oldengott:2017fhy,Choudhury:2020tka,Brinckmann:2020bcn,Das:2020xke,Mazumdar:2020ibx}, which similarly shrinks the neutrino free-streaming length. 
One may argue that the scalar becomes a dynamical field (acquiring the shift-symmetry breaking terms) only after a certain redshift, e.g., $z = 10^4$, such that the aggressive free-streaming limit does not apply in general. But there is no way to avoid the robust limit set at $z^{}_{\rm eq} = 3000$, because the CMB data require the dominance of matter component after that redshift.
One should be aware that a much more aggressive constraint can be put if the scalar-neutrino coupling is assumed to be switched on during the Big Bang nucleosynthesis~\cite{Venzor:2020ova}. We do not attempt to investigate this possibility in this work, hence we restrict ourselves to the scenario that the shift-symmetry of the scalar is broken after nucleosynthesis.

If these two criteria are satisfied, during the observable history of CMB the ultralight scalar will play the role of cold DM and neutrinos will stream as free particles.
We can evolve the current Universe back to higher redshift for the CMB constraints.
To be more specific, we require $g \phi < T^{}_{\nu}$ and $g^2 T^2_{\nu} < m^2_{\phi}$ before $z=3000$ for the conservative limit, and $g \phi < T^{}_{\nu} $ before $z=10^5$ for the aggressive one.
For the aggressive limit, if $g \phi < T^{}_{\nu} $ is required to be satisfied at exactly $z=10^5$  we find the condition  $g^2 T^2_{\nu} > m^2_{\phi}$ can also be met. According to our previous analysis, neutrinos will remain relativistic for $g^2 T^2_{\nu} > m^2_{\phi}$. Hence, the limit should be given at a lower redshift, such that there is no room for the parameter space to satisfy $g^2 T^2_{\nu} > m^2_{\phi}$. Such redshift is found to be $z=6 \times 10^4$.
We obtain our CMB limits as
\begin{eqnarray}\label{eq:conservative}
g & < & 0.4 \cdot \frac{m^{}_{\phi}}{\rm eV} \hspace{0.5cm} ({\rm conservative}) , \\
\label{eq:aggressive}
g & < & 0.09 \cdot \frac{m^{}_{\phi}}{\rm eV} \hspace{0.34cm} ({\rm aggressive}),
\end{eqnarray}
which correspond to the limits on the effective neutrino mass $g \phi$ in our local galaxy
\begin{eqnarray}\label{eq:}
m^{}_{\nu, {\rm eff}} & < & 9 \times 10^{-4}~{\rm eV} \hspace{0.4cm} ({\rm conservative}) , \\
m^{}_{\nu, {\rm eff}} & < & 2 \times 10^{-4}~{\rm eV} \hspace{0.4cm} ({\rm aggressive}).
\end{eqnarray}
These limits are dominated by the neutrino free-streaming requirement.
We need to point out that a thorough analysis of the CMB spectrum by simulating the evolution of perturbations in this context will make the results more robust, and we leave it as a possible future work.

\section{Implication for $\boldsymbol{0\nu\beta\beta}$ decays}\label{sec:0NuBB}
\noindent
There are many dedicated experiments that are looking for the signatures of $ 0\nu\beta\beta$ decays, e.g., GERDA 
Phase-II~\cite{Agostini:2018tnm}, CUORE \cite{Alduino:2017ehq}, SuperNEMO \cite{Barabash:2011aa}, KamLAND-Zen \cite{KamLAND-Zen:2016pfg} and EXO \cite{Agostini:2017jim}.
The discovery of lepton-number-violating processes in the future can reveal the Majorana nature of massive neutrinos~\cite{Dolinski:2019nrj}.

As has been shown in Fig.~\ref{fig:Feyn},
even without the vacuum Majorana neutrino mass the $0\nu\beta\beta$ transition can still take place in the background of ultralight DM.
The induced transition is very similar to the  Majoron emission process in vacuum.
However, because the vacuum in our case is occupied by dense coherent scalars, the dominant process will be the ``stimulated'' emission and absorption of the scalar particle, in analogy with the atomic transition in a laser field. 

Because the occupation number of scalars $N= n^{}_{\phi} \lambda^3$ is very large, such a transition will leave the DM field almost unaffected. 
This is equivalent to simply taking the expectation value for all relevant operators in the Lagrangian or scattering matrix.
In the presence of ultralight DM, the effective neutrino mass ($m_{ee}$) will hence receive an additional time-varying contribution
\begin{eqnarray}\label{eq:mbbt}
{m}^{\prime}_{ee}(t) &= & \langle \phi |  m^{}_{ee}+ g^{}_{ee} \hat{\phi}(t,x)|\phi \rangle  \notag\\
& = &  m^{}_{ee}+g^{}_{ee} {\phi}(t,x) \;,
\end{eqnarray}
%
where only the $g^{}_{ee}$ element of the general coupling matrix is relevant for $0\nu\beta\beta$ decays.
The $0\nu\beta\beta$-decay process typically takes place with a very small uncertainty of time.
%
It is therefore a good approximation to take the factor $\cos{ m^{}_{\phi} t}$ as a constant during the transition.
But the long-term events will feature a time-varying pattern if enough data have been collected.

The above result should be able to be reproduced from the more fundamental quantum field theory.
However, it is not apparent to obtain the time-varying effect from a view of the Feynman diagrams as illustrated in Fig.~\ref{fig:Feyn}, where the emission and absorption processes are themselves time-independent.
To understand this, we note that the mass of the ultralight DM ($< {\rm eV}$), namely the change in the energy of final electrons, is much smaller than the energy resolution of $0\nu\beta\beta$-decay experiments ($\sim {\rm keV}$). Hence, it is not possible at the quantum level to distinguish between the emission and absorption diagrams, which will lead to an interference. One can  find that it is this interference that gives rise to the time variation.
To be more precise, after the $0\nu\beta\beta$ transition via the vertex $(m^{}_{ee}+g \hat{\phi}) \overline{\nu^{\rm c}_{\rm L}} \nu^{}_{\rm L}/2$  the scalar state will be taken to 
\begin{eqnarray}\label{eq:}
& & \left[ m^{}_{ee}+g \hat{\phi}(t,x) \right]  \left|\phi \right\rangle  \notag \\
& & = \left[m^{}_{ee}+ g \sqrt{\frac{2}{m^{}_{\phi}V}} \left(\hat{a} \mathrm{e}^{-\mathrm{i} m^{}_{\phi}t} + \hat{a}^\dagger \mathrm{e}^{\mathrm{i} m^{}_{\phi}t}\right) \right]  |\phi \rangle \;.
\end{eqnarray}
Note that $|\phi \rangle$, $\hat{a}|\phi \rangle$ and $\hat{a}^\dagger|\phi \rangle$ almost completely overlap with each other, e.g., $\langle \phi | \hat{a}\, \hat{a} |\phi \rangle = N \mathrm{e}^{2\mathrm{i}\theta} $.
Together with other terms, this feature results in a probability proportional to $\Sigma^{}_{N} |\left\langle N \right|  m^{}_{ee}+g \hat{\phi}(t,x)|\phi \rangle|^2 $, namely
$\left| m^{}_{ee}+g {\phi}(t,x) \right|^2$, which is exactly the result obtained with the simple treatment of classical field in Eq.~(\ref{eq:mbbt}). This actually verifies why treating the ultralight DM as a classical field is appropriate for the quantum transition process.
The above observations are also applicable to the impact on neutrino oscillation experiments.

As for our $0\nu\beta\beta$ process, the half-life of a given isotope can be expressed as~\cite{Rodejohann:2011mu}
\begin{equation}
\dfrac{1}{T^{0\nu}_{1/2}} = G^{}_{0\nu} \cdot \left|{ M}^{}_{0\nu}\right|^{2} \cdot \frac{\left|m^{\prime}_{e e}(t) \right|^{2} }{M^2_e }\; ,
\label{eq:halflifedoublebeta}
\end{equation}
where  $  G_{0\nu}$ represents the phase-space factor, $ M_{0\nu} $ is the nuclear matrix element (NME), and $M^{}_{e} \approx 0.51~{\rm MeV}$ is the electron mass. 
The present lower limit on the half-life of $0\nu\beta\beta$ decays  from KamLAND-Zen collaboration~\cite{KamLAND-Zen:2016pfg} is $T^{0\nu}_{1/2} > 1.07 \times 10^{26}$ yr at 90\% C. L.
Using the phase-space factor $G^{}_{0\nu} = 3.793 \times 10^{-14}~{\rm yr^{-1}}$~\cite{Kotila:2012zza} with $g=1.27$ and NME $M^{}_{0\nu} \in (1.5 - 4.2)$~\cite{Guzowski:2015saa,Cirigliano:2018hja,Jokiniemi:2021qqv} for $^{136}{\rm Xe}$, the upper bound on the effective neutrino mass in standard three-flavor neutrino scenario is obtained as  $  | m_{ee}|  < (61 - 169)$ meV.
Here, $m^{}_{ee}$ is given by
%
\begin{eqnarray}\label{eq:mbb}
{m}^{}_{ee} &  \equiv & m^{}_1 \cos^2 \theta^{}_{13} \cos^2 \theta^{}_{12} e^{{\rm i}\rho} + m^{}_2 \cos^2 \theta^{}_{13} \sin^2 \theta^{}_{12}  \notag \\
& &+ m^{}_3 \sin^2 \theta^{}_{13} e^{{\rm i}\sigma} \; ,
\end{eqnarray}
where $m^{}_i$ (for $i = 1, 2, 3$) stand for the absolute masses of three neutrinos, $\{\theta^{}_{12},\theta^{}_{13} \}$ are the leptonic mixing angles, and $\{\rho, \sigma\}$ are the Majorana CP phases. 
%

\begin{figure}[t]
	\begin{center}
		\includegraphics[width=0.43\textwidth]{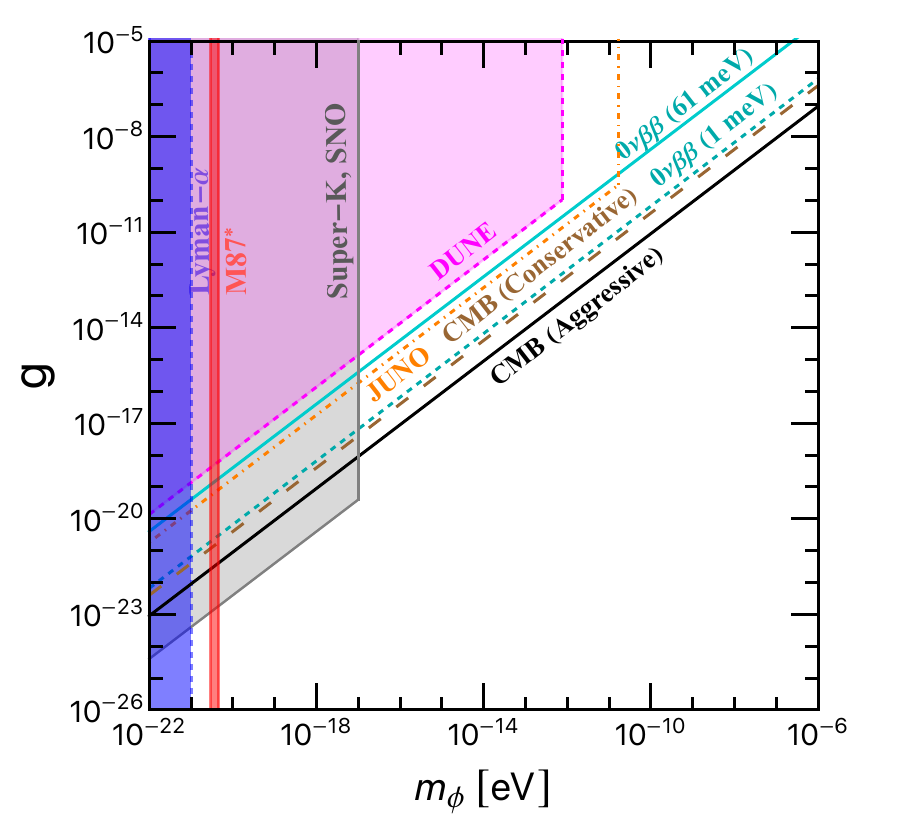}
	\end{center}
	\vspace{-0.3cm}
	\caption{\footnotesize Exclusion and sensitivities of various experiments on the parameter space of the ultralight DM mass $m^{}_{\phi}$ and the coupling to neutrinos $g$. The exclusion using the result of KamLAND-Zen are given by  cyan lines, while the CMB bounds in present work are shown as the dashed brown and solid black lines. See the text for details of other limits.}
	\label{fig:MphiGphi}
\end{figure}

To find out how the ultralight scalar exactly modifies the rate, we expand the vacuum and DM contributions
\begin{eqnarray}\label{eq:}
|{m}^{\prime}_{ee}|^2 & = & \left|m^{}_{ee}\right|^2 +  g^{2} \phi^2_0 \cos^2{m^{}_{\phi} t} \notag\\
& & + 2 |m^{}_{ee}| g^{} \phi_0 \cos\eta \cos{m^{}_{\phi}t }\; ,
\end{eqnarray}
where $\eta$ is the relative phase between $m^{}_{ee}$ and $g\phi$, the second term is induced solely by the DM field, and the third term is the interference with the vacuum neutrino mass.
Such time-varying effect is in principle observable, if enough $0\nu\beta\beta$-decay events with proper time resolution have been obtained.
When the ultralight DM oscillates fast enough, the long-term decay rate depends on the time average of $|{m}^{\prime}_{ee}|^2$, i.e.,
\begin{eqnarray}\label{eq:MeeApprox}
\overline{|{m}^{\prime}_{ee}|^2} = \left|m^{}_{ee}\right|^2 +  \frac{1}{2} g^{2} \phi^2_0 \;.
\end{eqnarray}
Note that the interference term has been averaged out in such case. 
The coupling constant $g^{}_{}$ is in general not related to $m^{}_{ee}$. Hence, when the cancellation of effective neutrino mass occurs, i.e., $m^{}_{ee} = 0$ for normal ordering, the DM-induced term can dominate the transition. 

 We present our noteworthy results as Fig.~\ref{fig:MphiGphi} in the $m^{}_{\phi}$-$g$ parameter plane. The exclusion regions of $ 0\nu\beta\beta $-decay experiments and cosmology are shown along with bounds and sensitivities arising from different experimental searches.
The vertical blue band shows the constraint based on the results of Lyman-$ \alpha $ forest. It has been pointed out (see for detailed analysis in Ref.~\cite{Hui:2016ltb}) that fuzzy DM lighter than $ \sim 10^{-21} $ eV is in  tension with observations of the Lyman-$ \alpha $ forest.
While the vertical thin-red band signifies the bound arising from the recent observations of the Event Horizon Telescope on M87$^*$~\cite{Davoudiasl:2019nlo}.
The gray shaded region gives
the limit from  Super-K and Sudbury Neutrino Observatory (SNO) neutrino oscillation experiments~\cite{Berlin:2016woy,Super-Kamiokande:2003snd,SNO:2005ftm}. 
The sensitivity from  forthcoming neutrino oscillation experiments~\cite{Krnjaic:2017zlz,Dev:2020kgz} like DUNE \cite{DUNE:2015lol} and JUNO \cite{JUNO:2015zny} have also been presented using the magenta shaded region and orange dot-dashed line, respectively. 
%

\begin{figure}[t]
	\begin{center}
		\includegraphics[width=0.43\textwidth]{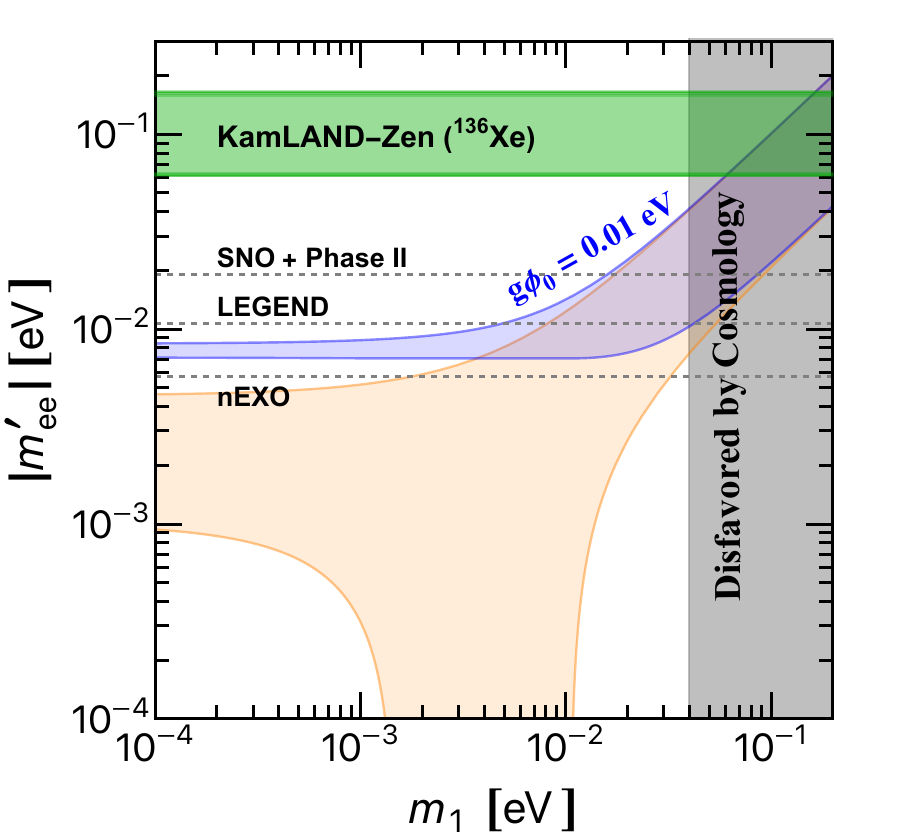}
	\end{center}
	\vspace{-0.3cm}
	\caption{\footnotesize The effective Majorana neutrino mass as a 
function of the lightest active neutrino mass $m_1$. 
The orange region stands for the standard case with normal mass ordering for neutrinos, while the blue region is for the case with $g\phi^{}_{0} = 0.01~{\rm eV}$.
}
	\label{fig:EffMee}
\end{figure}

The constraint from   $0\nu\beta\beta$-decay searches (KamLAND-Zen~\cite{KamLAND-Zen:2016pfg}) is shown as the solid cyan line. 
The projection with an ultimate sensitivity goal of $0\nu\beta\beta$-decay experiments $|m^{\prime}_{ee}| = 1~{\rm meV}$~\cite{Ge:2016tfx,Penedo:2018kpc,Cao:2019hli,Huang:2020mkz} is shown as the dotted cyan line.
In order to translate results into the $m^{}_{\phi}$-$g$ plane, we have used Eq.~(\ref{eq:MeeApprox}).
We present our  CMB bounds as the dashed brown  and solid black  lines, respectively. Two different limits corresponding to conservative and aggressive considerations are shown, as given by Eqs.~(\ref{eq:conservative}) and (\ref{eq:aggressive}), respectively. 
It can be noticed that JUNO can cover the limit of $ 0\nu\beta\beta $-decay experiment for most of the parameter spcae. However, for the region where the ultralight DM mass is greater than $10^{-11}~{\rm eV}$, $ 0\nu\beta\beta $-decay experiments set tighter bound than neutrino oscillation experiments.
On the other hand,  Super-K and SNO put the best constraint for ultralight DM below $ 10^{-17}~{\rm eV}$, whereas
the CMB analysis provides the most stringent limit for ultralight DM above $ 10^{-17} $ eV.

Finally, as a visual guide for the ultralight DM searches at $ 0\nu\beta\beta $-decay experiments, we present the effective Majorana neutrino mass, in presence of ultralight DM,  in Fig.~\ref{fig:EffMee}.  The horizontal green band indicates the current experimental limit from KamLAND-Zen, $(61 - 165)~{\rm meV}$~\cite{KamLAND-Zen:2016pfg}.  The dashed lines correspond to the future sensitivities projected  by SNO+ with $(19 - 46)~{\rm meV}$~\cite{Andringa:2015tza},  LEGEND with $(10.7 - 22.8)~{\rm meV}$~\cite{Abgrall:2017syy}, and nEXO with $(5.7 - 17.7)~{\rm meV}$~\cite{Albert:2017hjq}, respectively.
The vertical gray band represents the current neutrino mass limit of cosmological data from the Planck collaboration \cite{Aghanim:2018eyx}.
The orange region depicts the standard three-flavor neutrino scenario.
The impact of ultralight DM has been demonstrated using the blue region for a benchmark choice of $g\phi^{}_{0} = 0.01~{\rm eV}$.
It turns out that even the averaged effect of ultralight DM  can be tested by the next generation of $ 0\nu\beta\beta $-decay experiments. 
The presence of the coupling with ultralight DM can prevent $|m^{\prime}_{ee}|$ from falling into the well. Similar effect can also be induced by other new physics scenario as discussed in Ref.~\cite{Graf:2020cbf}. 

\section{Conclusion}\label{sec:conclusion}
\noindent
In this work, we investigate the impact of ultralight DM on $ 0\nu\beta\beta $-decay experiments. 
We find that the scalar DM can directly induce the $ 0\nu\beta\beta $-decay transition, even without the vacuum neutrino mass term.
The effect can be significant for very small couplings between the ultralight DM and neutrinos, due to the stimulated absorption and emission of scalars in the phase space filled with identical particles. As the consequence, the vacuum Majorana neutrino mass will be modified with a time-varying term.

We also explore the consequence of such a scalar field on cosmology. There are mainly two effects. First, the neutrino plasma will induce a screening mass to the ultralight DM, such that the DM field evolves as radiation instead of matter component. Second, the dense ultralight DM in the early Universe will shrink  the neutrino free-streaming length significantly and is in tension with the CMB observations.

The constraints and sensitivities of different experiments are given in Fig.~{\ref{fig:MphiGphi}}.
We conclude that $ 0\nu\beta\beta $-decay experiments can give stringent constraint compared to the forthcoming neutrino oscillation experiments like DUNE and JUNO for the DM mass above $\thicksim 10^{-11} $ eV.
In addition, the CMB observations have the best constraining capability for the mass range above $\thicksim 10^{-17} $ eV. 
Furthermore, we observe from Fig.~\ref{fig:EffMee} that our framework of ultralight DM can be tested even without analyzing the time-varying effect by the next generation of $ 0\nu\beta\beta $-decay experiments. 

\acknowledgments
\noindent
Authors are thankful to Werner  Rodejohann for his valuable comments and careful reading of the manuscript. Authors  gratefully acknowledge  Peter B. Denton for his suggestion to include the recent bounds from the measurement of M87$^*$.
Authors would also like to thank Eligio Lisi, Pablo Martínez-Miravé,  Manibrata Sen, and Shun Zhou for fruitful discussions. GYH is supported by the Alexander von Humboldt Foundation. NN is supported by the Istituto Nazionale di Fisica Nucleare (INFN) through the “Theoretical Astroparticle Physics” (TAsP) project.


\bibliographystyle{utcaps_mod}
\bibliography{references}

\end{document}